\documentclass[12pt]{article} 
 \usepackage{epsfig}
 \usepackage{graphicx}
 \usepackage{pict2e}
\newcommand{\sfrac}[2]{{\textstyle\frac{#1}{#2}}}
\renewcommand{\SS}{\mathcal{S}}
\newcommand{\eps}{\varepsilon}
\begin{document} 
\title{A spatial model of city growth and formation} 
\author{D J Aldous\thanks{Department of Statistics, 367 Evans Hall \#\  3860,
 U.C. Berkeley CA 94720;  aldous@stat.berkeley.edu;
 www.stat.berkeley.edu/users/aldous.  Aldous's research supported by
N.S.F Grant DMS-0704159. } 
\and  B  Huang
}

\date{}
 \maketitle

\begin{abstract} 
We introduce a model in which city populations grow at rates proportional to the area of their ``sphere of influence", where the influence of a city depends on its 
population (to power $\alpha$) and distance from city (to power $- \beta$) and where new cities arise according to a  certain random rule.  
A simple non-rigorous analysis of asymptotics indicates that for $\beta > 2 \alpha$ the system exhibits ``balanced growth" in which there are an increasing number of large cities, 
whose populations have the same order of magnitude, whereas for  $\beta < 2 \alpha$ the system exhibits ``unbalanced growth" in which a few cities capture most of the total population. 
Conceptually the model is best regarded as a spatial analog of the combinatorial  ``Chinese restaurant process".
\end{abstract} 



\maketitle

\section{Introduction}
\label{sec-intro}

There is substantial literature in Economics concerning locations and population sizes of cities,  a central quantitative feature of the latter being
 the observation ({\em  Zipf's law}) that the number of cities with populations larger than $s$ scales roughly as $s^{-1}$.
 Useful background can be found in the 2004 survey \cite{survey} which describes
``both bare-bone statistical theories and more developed economic theories."  
The former, exemplified by the Gibrat model (proportional growth rates of cities are random but independent of population size) 
might better be called purely mathematical models, while the latter (quoting \cite{survey})
``reflect such important economic forces as increasing returns, congestion, trade and non-market interactions". 
But (as in the broader literature on spatial economics featured in the monograph \cite{krugman}) most models are 
not truly ``spatial" in the sense that the geometry of two-dimensional space plays an essential role.
The purpose of this paper is to present a purely mathematical model which is explicitly spatial in this sense.
The model is not intended as literally realistic for cities, but rather as a novel style of model (see discussion of related models in section \ref{sec-related}) and one for which (unlike many explicitly spatial models in other contexts) 
 non-obvious properties can be derived via quite simple albeit non-rigorous arguments.

\section{The model}
At each time step
$t = 1,2,3,\ldots$ 
there are cities at positions $x_i$ in the unit square
$[0,1]^2$, 
with populations $N_i(t) \geq 1$, 
the total population being 
$\sum_i N_i(t) = t$.
The model has three parameters
\[ 0 < c_0 < \infty, \quad 0 < \alpha \le 1, \quad \beta > 0  \]
which are used to define a function
\begin{equation} I_0(n,r) = c_0 n^\alpha r^{-\beta} \label{I0}
\end{equation}
interpreted as the ``influence" of a city of population $n$ at a point at distance $r$ 
from the city.
For a position 
$y \in [0,1]^2$ define
\begin{equation}
I(y,t) = \max_i I_0(N_i(t),|y-x_i|) 
= c_0 \max_i N_i^\alpha(t) |y-x_i|^{-\beta} 
\label{def-I}
\end{equation} 
(the maximum influence at that position) 
and then define the 
{\em sphere of influence} 
of city $i$
to be the region 
\begin{equation}
\SS(i,t) = \{y: I_0(N_i(t),|y-x_i|) = I(y,t) \}
\label{def-SS}
\end{equation} 
in which city $i$ has larger influence than any other city.
At time $1$ there is a
single city of population $1$
at a uniform random point of
$[0,1]^2$.
The general evolution rule is:
\begin{quote}
At time $t+1$ an immigrant arrives at a uniform random position
$U$ in $[0,1]^2$,
and either 
\\
(i) (with probability $1/(1+ I(U,t))$) 
founds a new city at position $U$ with population $1$;\\
or (ii) (with probability $I(U,t)/(1+ I(U,t))$)
joins the city $i$ whose sphere of influence contains $U$, 
thereby increasing its population to $N_i(t+1) = N_i(t) + 1$.
\end{quote}

\subsection{Remarks on the model}
{\bf 1.} If city populations were equal then the partition into spheres of influence would be just the usual
Voronoi tessellation  \cite{1770006}; in general one can consider it as a form of weighted 
Voronoi tessellation.

\noindent
{\bf 2.} 
The two qualitative features of the model are \\
(i) the growth rate of a city depends on its size and on the sizes and distances of other cities\\
(ii) a certain stochastic rule for founding of new cities.\\ 
One could imagine many different rules to formalize these features;
while there is no necessary connection between the two features,
our formulation in which both are derived via the same influence function is mathematically convenient.

\noindent
{\bf 3.} Given the configuration at a large time $t$, the subsequent evolution over a relatively small time interval is deterministic to first order, because 
a city population grows at rate proportion to the area of its sphere of influence. 
Randomness plays a role both via a ``founder effect" (the random positions of the first few cities) and more subtly, in the 
 ``balanced growth" case, because the newly-founded cities at (non-uniform) random positions grow comparatively rapidly to attain the same order of magnitude population as the older cities.

\noindent
{\bf 4.} The parameter $c_0$ has a quantitative influence via the founder effect but does not affect the types of asymptotic behavior we discuss; the model has the two essential parameters 
$\alpha$ and $\beta$ which do affect this behavior.

\noindent
{\bf 5.} The case $\alpha = 1$ is conceptually closest to previous models (see section \ref{sec-related}) and seems worthy of more detailed study. 
The case $\alpha > 1$ is less interesting because one gets explosive growth without considering any spatial interaction.

\noindent
{\bf 6.} We modeled population growth as via single ``immigrants" for simplicity -- more elaborate models with population growth caused by a surplus of births over deaths can be expected to exhibit similar behavior.

\section{Analysis of long-time behavior} 
We first consider the case  $0<\alpha < 1$.
We can analyze quantitatively the growth exponents of several quantities, implicitly assuming certain qualitative behavior discussed below.
The quantities we study are

$N^*(t) = $ typical city population

 $R^*(t ) = $ distance from from typical point to nearest city 
 
 $I^*(t) = $ value of the influence function $I(y,t)$ at a typical point $y$.
 
 \noindent
Write $M(t)$ for the number of cities at time $t$, and suppose their populations are mostly the same order of magnitude.   
 Clearly
\[ N^*(t) \approx t/M(t); \quad  R^*(t)  \approx M^{-1/2}(t) \]
and this implies
\[ I(y,t) \approx (N^*(t))^\alpha (R^*(t))^{-\beta} \approx
(\frac{t}{M(t)})^\alpha \ M^{\beta/2}(t)  \approx
t^\alpha M^{-\alpha + \beta/2}(t) . \]
The probability that a new arrival founds a new city is  $\approx 1/I^*(t)$, so 
we get an equation
\begin{equation}
 \sfrac{dM}{dt} \approx \sfrac{1}{I^*(t)} \approx t^{-\alpha} M^{-\beta/2 + \alpha} . 
 \label{eq-heuristic}
 \end{equation}
This has solution
\[ M(t) \approx t^\theta, 
\quad \mbox{ for } \theta = \frac{1-\alpha}{1-\alpha + \beta/2} \]
obtained from solving 
$\theta - 1 = - \alpha + \theta(\alpha - \beta/2)$.  
Note that the typical influence is therefore 
\begin{equation}
 I^*(t) \approx (dM(t)/dt)^{-1} 
\approx t^{1 - \theta} ; \quad 
 1 - \theta = \sfrac{\beta}{2 - 2 \alpha + \beta} 
 \label{heur-I}
 \end{equation} 
 and the typical distance to nearest city is 
 \begin{equation}
  R^*(t) \approx M^{-1/2}(t) \approx t^{-\theta/2}; \quad 
  \theta/2 = \sfrac{1-\alpha}{2 - 2\alpha + \beta} 
  \label{heur-r}
  \end{equation}
  and the typical city population size is 
  \begin{equation}
   N^*(t) \approx \sfrac{t}{M(t)} \approx t^{1 - \theta}  ; \quad 
 1 - \theta = \sfrac{\beta}{2 - 2 \alpha + \beta} .
 \label{heur-N}
 \end{equation}

Now the calculations above  rest upon an intuitive picture of the qualitative behavior of the process, 
 that for large $t$ and a typical position $y$\\
 (a) most different cities' populations are the same order of magnitude \\
 (b) $y$ is in the sphere of influence of some {\em nearby} city \\
 (c) a city newly founded at $t$ will grow, in time $\delta t$, to some population which is $\eps(\delta)$ times the typical time-$t$ city population.
 
 \noindent
 Call this the {\em balanced growth scenario}.  
 But one can imagine an alternative picture, the 
 {\em unbalanced growth scenario},
 in which, for large $t$ and a typical position $y$ \\
 (d) $y$ is in the sphere of influence of some city $A$ at distance $r$ which is much larger than the distance to nearby cities\\
 (e) the nearby cities' populations are a smaller order of magnitude than city $A$'s, 
 and their spheres of influence are surrounded by that of city $A$\\
 (f) New cities grow extremely slowly.

To investigate these scenarios we use a self-consistency calculation.  
Consider a city founded at time $t$, and consider 
\[ \mbox{ $N(s) = $ population of this city at time $s$ after founding,} \] 
 looked at  over a relatively short time period $0<s<\frac{1}{100}t$, say. 
The radius $r(s)$ of its sphere of influence satisfies 
\[ N^\alpha(s) r^{-\beta}(s) \approx I^*(t) \approx t^{1 - \theta} .\]
The rate of population growth is proportional to 
area of sphere of influence, so we get the equation
\begin{equation}
 \frac{dN(s)}{ds}
\approx r^2(s)
\approx t^{-2(1-\theta)/\beta} \ 
N^{2\alpha/\beta}(s)
; \quad N(0) = 1 . 
\label{xyz}
\end{equation}
We now have two cases.

{\bf Case 1.}
$\beta < 2\alpha$.
Here the solution of 
$dy(s)/ds = y^{2\alpha/\beta}(s)$ 
explodes in finite time $s$, but stays bounded for some small time. 
So the solution $N(s)$ of (\ref{xyz})  stays bounded for some time $s$ of order 
$t^{2 (1-\theta)/\beta}$.
But the assumption 
$\beta < 2\alpha$ 
implies 
$2(1-\theta)/\beta
= \frac{1}{1-\alpha + \beta/2} 
> 1$ 
implying that 
$N(\sfrac{1}{100}t)$ is bounded, 
in contradiction to behavior (c) above.

{\bf Case 2.}
$\beta > 2\alpha$.
Here the solution of (\ref{xyz}) is
\[ N(s) \approx t^\xi (s + t^{-\xi/\phi})^\phi; \quad \mbox{ where } 
\phi = \frac{\beta}{\beta - 2 \alpha}, \quad 
\xi = \frac{2(\theta - 1)}{\beta - 2 \alpha} . \] 
Here $-\xi/\phi$ works out to be $\frac{1}{1-\alpha + \beta/2} < 1$ and
so $N(\sfrac{1}{100}t)$ is order $t^{\xi + \phi}$. 
A calculation shows $\xi + \phi = 1 - \theta$,
consistent with behavior (c) above.

\paragraph{Conclusion of the analysis.}  The self-consistency check provides convincing 
evidence for the conclusion
\begin{quote}
for $\beta > 2 \alpha$, the balanced growth scenario holds, with growth exponents given by (\ref{heur-I} - \ref{heur-N}).
\end{quote}
This cannot be true in the other case, so we predict the natural alternative qualitative behavior 
\begin{quote}
for $\beta < 2 \alpha$, the unbalanced growth scenario holds, with growth exponents
\begin{equation}
N^*(t) = t^{1 - o(1)}; \quad 
R^*(t) = t^{-o(1)}; \quad 
I^*(t) = t^{1 - o(1)} . 
\label{NRI-unb}
\end{equation}
\end{quote}
and one can give analogous self-consistency arguments for this case.
Note that these exponents are therefore discontinuous as $(\alpha,\beta)$ cross the boundary between the balanced and unbalanced regions.

In fact one can now {\em a posteriori} see a conceptually simpler distinction between the two scenarios.
Consider a city founded at time $t$.  
If the area of its sphere of influence upon founding is 
$> 1/t$ then its initial growth rate (proportional to size) will be larger
than the average growth rate of other cities, while if this area is smaller than $1/t$ its initial growth rate will be slower.  
This is the distinction between (c) and (f).
But to calculate this initial area in terms of $\alpha$ and $\beta$ one needs to go through the same calculations as before -- 
we do not see any simpler argument that these alternatives correspond to
$\beta > 2\alpha$ and $\beta < 2 \alpha$.

\subsection{The case $\alpha = 1$.}
The arguments above hold for $\alpha = 1$, but here the distinction between the two 
cases ($\beta < 2 \alpha$ or $\beta > 2 \alpha$) disappears, in that the predictions (\ref{heur-I} - \ref{heur-N})
and (\ref{NRI-unb}) of the two cases are the same.  
For this case $\alpha = 1$ we expect the number of cities to grow as some power of 
$\log t$, but we do not have any convincing argument for how this rate depends on $\beta$.  
Note that $\alpha = 1$ is the case where proportional growth rates do not depend on city size, as in the (non-spatial)  Gibrat model \cite{survey}  
often invoked to explain Zipf's law.
As observed in section \ref{sec-related} 
this case is loosely analogous to other models 
and perhaps the main contribution of this paper is to spotlight the case $\alpha = 1$
as a topic for more detailed future study.

\subsection{Simulation results}
We show simulations in the balanced growth scenario.
Figure 1 shows  city positions and sizes (indicated by the volume of the cubes) in a simulation with $\alpha = 0.2, \beta = 4.8$ and total 
population 300.  This is 
visually consistent with the qualitative behavior described earlier.

\begin{center} 
 \includegraphics[width=81mm]{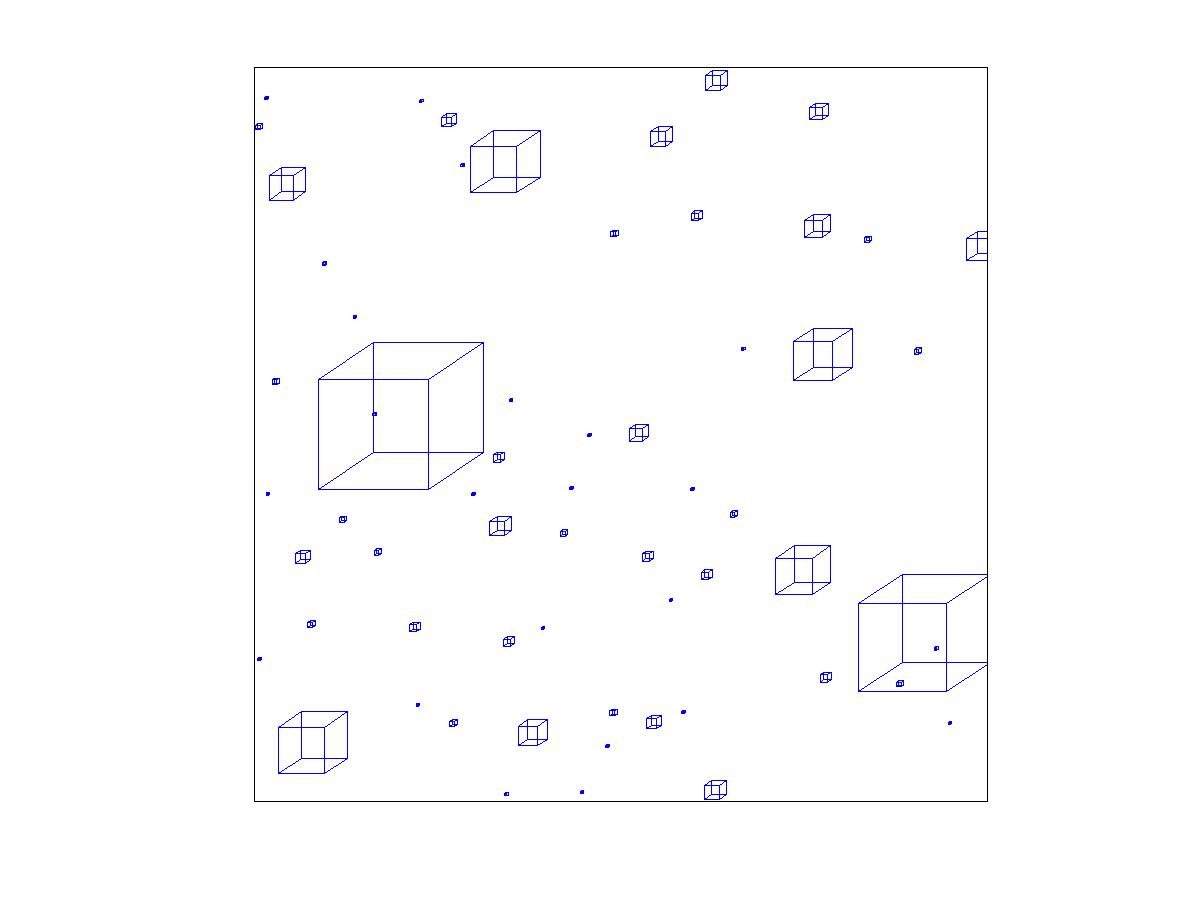}
\end{center}

{\bf Figure 1.}  City positions and sizes in a simulation of the balanced growth scenario.

\medskip
\noindent
Figure 2 shows results from simulations with 
$\alpha = 0.2$ and three values of $\beta$ chosen to make 
$\theta = 0.25, 0.5, 0.75$.  
The jagged lines are the simulation results and the straight lines have the slopes predicted by (\ref{heur-I} - \ref{heur-N}).

\begin{center} 
 \includegraphics[width=97mm]{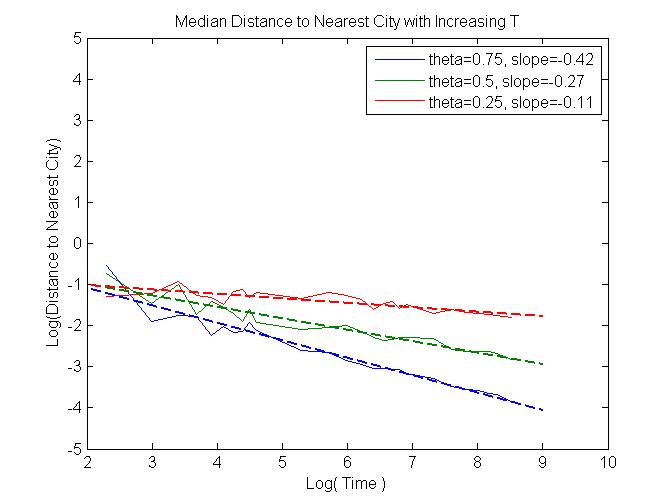}
 \includegraphics[width=97mm]{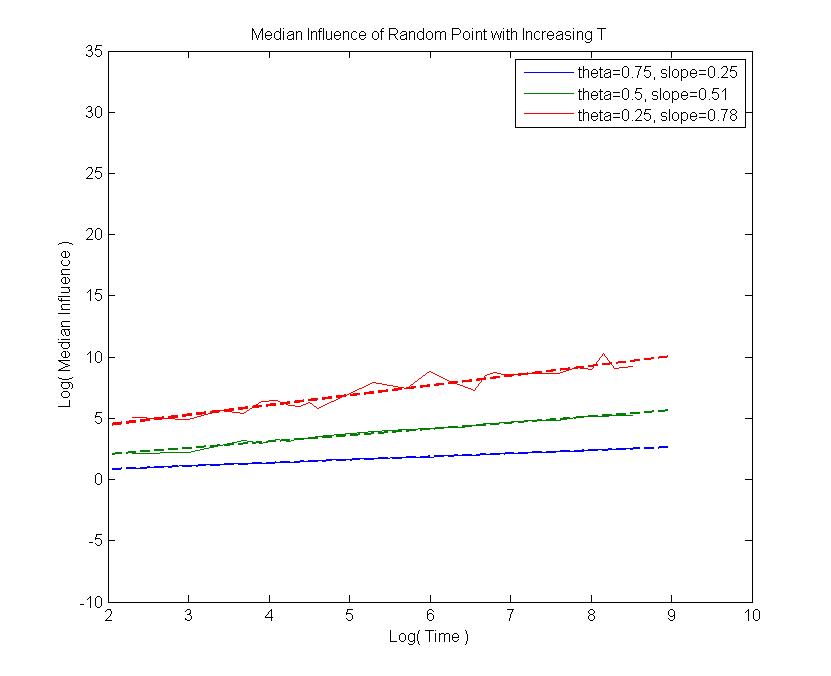}
 \includegraphics[width=97mm]{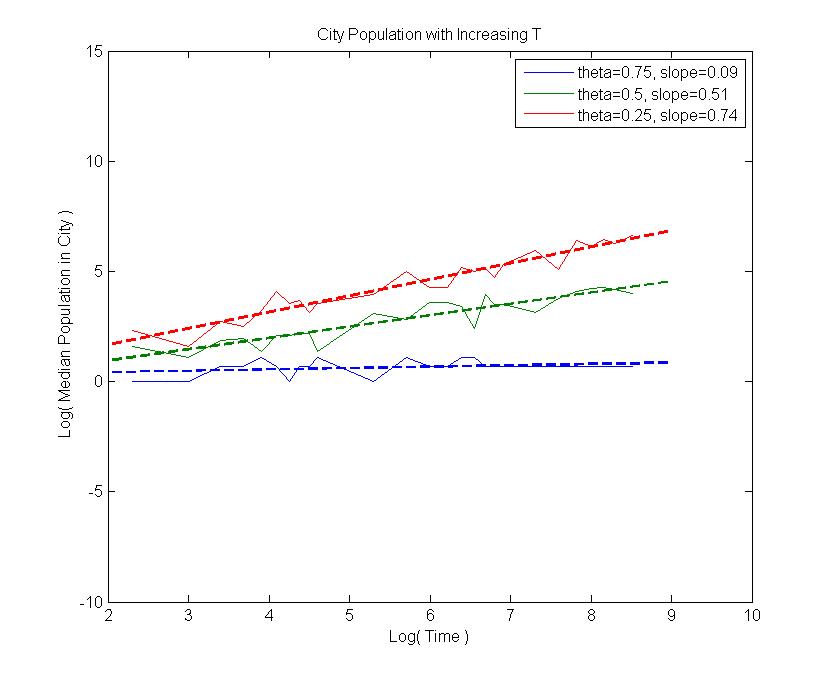}
 \end{center}

{\bf Figure 2.}  Simulation data fits the predicted power laws.

\section{Related models}
\label{sec-related}
We do not know any previous models that are closely related to ours. 
Amongst numerous distantly related models that have appeared in different disciplines within the mathematical sciences, let us mention four.

\paragraph{The Chinese restaurant process.}
In our terminology, this is the process where the arrival at time $t+1$ either \\
(i) (with probability $c_0/(t + c_0)$) 
founds a new city with population $1$;\\
or (ii) (with probability $N_i(t)/(t + c_0)$)
joins  city $i$. \\
See \cite{MR2245368} for a treatment of this model and some generalizations; these do not involve any spatial structure.  
A key feature of this model is that, for ordered city sizes 
$N_{(1)}(t) \ge N_{(2)}(t) \ge \ldots$, there is a limit distribution after 
normalizing by total population $t$:
\begin{equation}
 t^{-1}(N_{(1)}(t),N_{(2)}(t),\ldots) \to (X_1,X_2,\ldots ) , \quad 
\mbox{ where } X_i > 0, \ \sum_i X_i = 1
 \label{PD}
 \end{equation}
and the limit is the Poisson-Dirichlet distribution.  
The $\alpha = 1$ case of our model is a spatial analog, so it is natural to ask whether it has the same behavior (\ref{PD}), after appropriate normalization.  If so, then one can ask whether these limit sizes for large cities have power law distribution (as in Zipf's law) or a geometrically decreasing distribution (as in  Poisson-Dirichlet).  But such questions seem 
currently out of reach of analytic arguments.

Note that after 
originating in probabilistic combinatorics and mathematical genetics, 
the Chinese restaurant process and variants have found extensive use 
as general-purpose Bayes priors for statistical 
problems involving groups of data \cite{MR2279480}, 
 so it is not inconceivable that variants of our model would make useful priors for explicitly spatial data.

\paragraph{Coagulation models.}
There is a large literature in physical chemistry on {\em coagulation}, meaning coalescence of clusters of mass.  
Though the underlying picture is of motion in space (with coalescence when clusters meet), the usual models \cite{BC90} ignore spatial position and study deterministic equations for the density $f_i(t)$ of mass-$i$ clusters at time $t$; a parameter in the equations is a kernel $K(i,j)$ giving the propensity for mass-$i$ and mass-$j$ clusters to merge.  
Closest to our model is the special case of the 
Becker-D\"{o}ring equations 
\cite{king-wattis} 
of polymers growing by collisions with monomers; mass-$i$ clusters can grow only by coalescing with mass-$1$ clusters.

\paragraph{Random tessellations.}  
Turning to explicitly spatial models, within the discipline of 
{\it stochastic geometry} there are many models for random partitions of the plane, for instance 
random Johnson-Mehl tessellations \cite{MR199212}.
But we do not know models where such tessellations evolve by  stochastic dynamics comparable to our model.

\paragraph{A spatial network model.}  
A spatial analog of the popular ``proportional attachment" network models was studied by simulation in
\cite{barrat-05}.  This model has additional graph structure, but 
(interpreting their ``number of edges" as ``population") is essentially the following model.  
Take an integer parameter $m \geq 1$ and a ``distance scale" parameter $r_c$.

\noindent
(i) A city arrives at a uniform random point $y$ in a given region, and is given population $m$. \\
(ii) Simultaneously $m$ existing cities have their population increased by $1$, with city $i$ chosen with probability proportional to 
$N_i  \exp(- |y-x_i|/r_c)$.

\noindent  
This has similar ingredients to our model, but
their conclusions focus on network traffic properties, and so are not comparable to ours.


\end{document}